\documentclass{llncs}
\usepackage{html}

\title{ClassdescMP: Easy MPI programming in C++}

\author{Russell K. Standish\inst{1}\inst{2} and Duraid Madina\inst{2}}
\institute{School of Mathematics \and High Performance Computing Support Unit\\
University of New South Wales,Sydney, 2052,Australia\\
\email{\{R.Standish,duraid\}@unsw.edu.au}\\http://parallel.hpc.unsw.edu.au/classdesc}

\newcommand{\EcoLab}{{\sffamily\slshape
    \mbox{\raisebox{.5ex}{Eco}\hspace{-.4em}{\makebox[.5em]{L}ab}}}}

\begin{document}
\maketitle

\begin{abstract}
ClassdescMP is a distributed memory parallel programming system for
use with C++ and MPI. It uses the Classdesc reflection system to ease
the task of building complicated messages to be sent between
processes. It doesn't hide the underlying MPI API, so it is an
augmentation of MPI capabilities. Users can still call standard MPI
function calls if needed for performance reasons.
\end{abstract}

\section{Classdesc Reflection}

MPI is an {\em application programming interface} (API - in other
words library of functions calls) for passing data from one unix
process to another. The processes may be running on the same computer,
or completely distinct computers, so this system provides a means of
implementing {\em distributed memory parallel processing}.

MPI has been used to great success in a variety of Engineering and
Scientific codes where large arrays of the same type of data (eg floating
point numbers) need to be exchanged between processes. However, with object
oriented codes, one really needs to send {\em objects} (which may well
be compound) between processes. MPI does provide the \verb+MPI_Type+
functionality, which allows the description of compound structures to
be built up, however this is difficult to use, and doesn't handle the
case where the structure contains references to other objects (eg pointers).

Object {\em reflection} allows straightforward implementation of
serialisation (i.e. the creation of binary data representing objects that can
be stored and later reconstructed), binding of scripting
languages or GUI objects to `worker' objects and remote method
invocation.  Serialisation, for example, requires knowledge of the
detailed structure of the object. The member objects may be able to
serialised (eg a dynamic array structure), but be implemented in terms
of a pointer to a heap object. Also, one may be interested in
serialising the object in a machine independent way, which requires
knowledge of whether a particular bitfield is an integer or floating
point variable.

Languages such as Objective C give objects reflection by creating
class objects and implicitly including an {\em isa} pointer in objects
of that class pointing to the class object.  Java does much the same thing,
providing all objects with the native (i.e. non-Java) method {\tt getClass()}
which returns the object's class at runtime, as maintained by the virtual
machine.

When using C++, on the other hand, at compile time most of the
information about what exactly objects are is discarded. Standard C++
does provide a run-time type enquiry mechanism, however this is only
required to return a unique signature for each type used in the
program. Not only is this signature be compiler dependent, it could be
implemented by the compiler enumerating all types used in a particular
compilation, and so the signature would differ from program to
program!

The solution to this problem lies (as it must) outside the C++
language per se, in the form of a separate program
(classdesc)\cite{Madina-Standish01,Standish01b} which parses an input
program and emits function declarations that know about the structure
of the objects inside. These are generically termed {\em class
descriptors}. The class descriptor generator only needs to handle
class, struct and union definitions. Anonymous structs used in
typedefs are parsed as well.  What is emitted in the object descriptor
is a sequence of function calls for each base class and member, just
as the compiler generated constructors and destructors are
formed. Function overloading ensures that the correct sequence of
actions is generated at compile time.

Once a class definition has been parsed by classdesc, and the emitted
class descriptors included into the original program, an object of
that class can be serialised into a buffer object using the \verb+<<+
operator. The \verb+>>+ operator can be used to extract the value
back.
\begin{verbatim}
#include "foo.h"  //foo class definition
#include "foo.cd" //foo class descriptor
...
pack_t buf;
foo a=...;
bar b=...;
foo c;
buf << a << b;  // serialise a into buffer.
buf >> c;  // unpack buffer into c. Now c == a.
\end{verbatim}

Once a buffer is packed, it can be saved to a file, or transferred
over a network connection by using its public members \verb+data+,
which points to the data, and \verb+size+ which gives the number of
bytes currently in the buffer.

A variant of \verb+pack_t+ is the \verb+xdr_pack+ type, which uses the
XDR library to represent the objects in a machine independent
form. This is needed if the receiver of the serialised object is of a
different architecture (eg endianess, or wordsize) to the sender.

The only real reason for using \verb+MPI_Type+ in MPI is to support
messages of compound objects between differing processors in a
heterogenous cluster. \verb+xdr_pack+ gives heterogenous cluster
support ``for free''.

\section{MPIbuf}

\verb+MPIbuf+ is a derived type which inherits from \verb+pack_t+ (or
\verb+xdr_pack+ if the \verb+HETERO+ flag is set). Similar to the
notion of {\em manipulators} with the standard iostream library,
\verb+MPIbuf+ has manipulators, such as \verb+send()+:
\begin{verbatim}
buf << a << b << send(1);
\end{verbatim}
which sends {\tt a} and {\tt b} to process 1.

Process 1 can receive this message with:
\begin{verbatim}
buf.get() >> a >> b;
\end{verbatim}

Other manipulators and methods of MPIbuf implement collective
operations such as broadcast, gather and scatter:
\begin{verbatim}
buf << a << b << bcast(0) >> a >> b;
\end{verbatim}
suffices to broadcast the values of {\tt a} and {\tt b} to all
processors.

By default, an \verb+MPIbuf+ uses the \verb+MPI_COMM_WORLD+
communicator, which includes all processes started by the MPI
system. Collective operations can be restricted to subsets of
processes by defining an MPI {\em communicator} object using
\verb+MPI_Comm_create+ and assigned to the \verb+Communicator+ member
of the \verb+MPIbuf+ object.

\section{MPISPMD}

An MPI program requires a certain amount of structure to just to setup
the processes, and tear them down at the finish, before any real
coding starts.

A typical MPI program might do the following:
\begin{verbatim}
main(int argc,char**argv)
{
  MPI_Init(&argc,&argv);
  MPI_Comm_size(&nprocs);
  MPI_Comm_rank(&myid);
  ... computation ...
  MPI_Finalize();
}
\end{verbatim}

One problem that frequently occurs is forgetting to call
\verb+MPI_Finalize()+, which needs to be called from all
processes. This may occur because of an error condition, or otherwise
abnormal exit from the program. This is a problem, as typical MPI
implementations have resources such as shared memory segments that
need freeing, and the slave processes need to be terminated. On a
shared high performance computing system, stray processes consume CPU
resources that may be better employed by other users.

\verb+MPISPMD+ arranges for \verb+MPI_Finalize()+ to be called from its
destructor. This has two advantages: \verb+MPI_Finalize()+ will be
called as soon as the \verb+MPISPMD+ leaves its scope. Also, the
destructor will be called if an exception is raised, leading to
cleaner error handling. Unfortunately, if the program terminates from
a conventional low level call to \verb+abort()+, the cleanup will not
take place.

The above piece of code can be replaced by:
\begin{verbatim}
main(int argc, char** argv)
{
  MPISPMD C(argc,argv);
  ... computation ...
}
\end{verbatim}

The number of processes and processor ID can be queried from
MPISPMD::nprocs and MPISPMD::myid respectively. Note that you should
not declare the MPISPMD object as a global variable, as some MPI
implementations object to \verb+MPI_Finalize()+ being called after
\verb+main()+ exits. If you need to refer to the MPISPMD object from
global scope, declare a global pointer to MPISPMD, and initialise this
from with \verb+main()+:
\begin{verbatim}
MPISPMD *C;
main(int argc, char** argv)
{
  MPISPMD comp(argc,argv); C=&comp;
  ...
\end{verbatim}

\section{MPIslave}

MPISPMD gives rather minimal support, as the SPMD programming model is
already implicit in MPI. Another programming model which MPI is often
put to is master-slave processing. Setting up the structure of a
master-slave program is very tedious and error prone. The
\verb+MPIslave+ class is designed to make master-slave algorithms
simple to program.

When a \verb+MPIslave+  object is instantiated, a slave
``interpreter'' object is instantiated on each process to receive
messages from the master. As \verb+MPIslave+ needs to know the type of
object to be instantiated on the slave processes, it is implemented as
a template, with the type of slave object passed as the template
parameter.

A message sent to the slave process starts with a method pointer of
type: \verb+void (S::*)(MPIbuf&)+ where \verb+S+ is the slave object
type, followed by the arguments to be passed. That method of the slave
object is then called, with the arguments passed through \verb+MPIbuf+
argument, and any return values also passed through \verb+MPIbuf+
argument:
\begin{verbatim}
struct S
{
  void foo(MPIbuf& args)
  {
    int x,y,r;
    args >> x >> y;
    ...
    args.reset() << r;
  }
};

main(int argc, char** argv)
{ 
  MPIslave<S> C;
  MPIbuf buf;
  int x=1, y=2;
  buf << &S::foo << x << y << send(1);
}
\end{verbatim}

When the \verb+MPIslave+ object is destroyed on the master process, it
arranges for all the slave objects to be \verb+MPI_Finalized()+ and
destroyed also.

\verb+MPIslave+ also has features for managing a pool of idle slaves:
\begin{verbatim}
MPIslave<S> C(argc,argv);
vector<job> joblist;
for (int p=1; p<C.nprocs && p<joblist.size(); p++)
   C.exec(C << &S::do_job << joblist[p]);
while (p<joblist.size())
  {
    process_return(C.get_returnv());
    C.exec(C << &S::do_job << joblist[p++]);
  }
while (!C.all_idle()) 
  process_return(C.get_returnv());
\end{verbatim}

\section{Performance}

Serialisation within classdesc is very efficient, as the relevant
class descriptor functions are all inlined. The performance is
comparable with that of simply copying the data into the buffer
object. Using the MPIbuf mechanism incurs the overhead of copying
into, and back out of an extra buffer, which is particularly wasteful
if only one object is being sent. Clearly, within performance critical
code, Classdesc MPIBuf operations may be replaced manually by an
equivalent sequence of \verb+MPI_Send()+s and \verb+MPI_Recv()+s. As
with any optimisation technique, this should be performed after
profiling the code to ensure benefit.

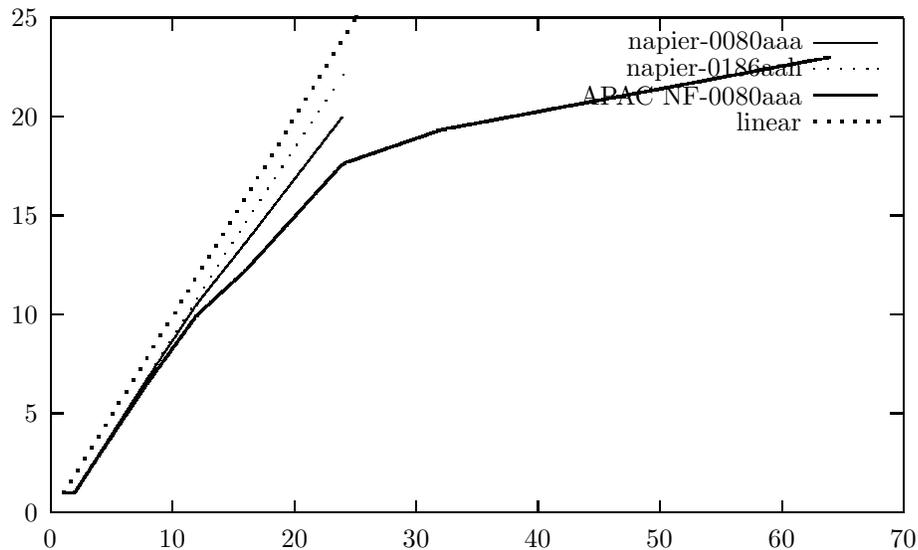
\begin{figure}
\setlength{\unitlength}{0.240900pt}
\ifx\plotpoint\undefined\newsavebox{\plotpoint}\fi
\sbox{\plotpoint}{\rule[-0.200pt]{0.400pt}{0.400pt}}%
\begin{picture}(1500,900)(0,0)
\font\gnuplot=cmr10 at 10pt
\gnuplot
\sbox{\plotpoint}{\rule[-0.200pt]{0.400pt}{0.400pt}}%
\put(100.0,82.0){\rule[-0.200pt]{4.818pt}{0.400pt}}
\put(80,82){\makebox(0,0)[r]{0}}
\put(1419.0,82.0){\rule[-0.200pt]{4.818pt}{0.400pt}}
\put(100.0,238.0){\rule[-0.200pt]{4.818pt}{0.400pt}}
\put(80,238){\makebox(0,0)[r]{5}}
\put(1419.0,238.0){\rule[-0.200pt]{4.818pt}{0.400pt}}
\put(100.0,393.0){\rule[-0.200pt]{4.818pt}{0.400pt}}
\put(80,393){\makebox(0,0)[r]{10}}
\put(1419.0,393.0){\rule[-0.200pt]{4.818pt}{0.400pt}}
\put(100.0,549.0){\rule[-0.200pt]{4.818pt}{0.400pt}}
\put(80,549){\makebox(0,0)[r]{15}}
\put(1419.0,549.0){\rule[-0.200pt]{4.818pt}{0.400pt}}
\put(100.0,704.0){\rule[-0.200pt]{4.818pt}{0.400pt}}
\put(80,704){\makebox(0,0)[r]{20}}
\put(1419.0,704.0){\rule[-0.200pt]{4.818pt}{0.400pt}}
\put(100.0,860.0){\rule[-0.200pt]{4.818pt}{0.400pt}}
\put(80,860){\makebox(0,0)[r]{25}}
\put(1419.0,860.0){\rule[-0.200pt]{4.818pt}{0.400pt}}
\put(100.0,82.0){\rule[-0.200pt]{0.400pt}{4.818pt}}
\put(100,41){\makebox(0,0){0}}
\put(100.0,840.0){\rule[-0.200pt]{0.400pt}{4.818pt}}
\put(291.0,82.0){\rule[-0.200pt]{0.400pt}{4.818pt}}
\put(291,41){\makebox(0,0){10}}
\put(291.0,840.0){\rule[-0.200pt]{0.400pt}{4.818pt}}
\put(483.0,82.0){\rule[-0.200pt]{0.400pt}{4.818pt}}
\put(483,41){\makebox(0,0){20}}
\put(483.0,840.0){\rule[-0.200pt]{0.400pt}{4.818pt}}
\put(674.0,82.0){\rule[-0.200pt]{0.400pt}{4.818pt}}
\put(674,41){\makebox(0,0){30}}
\put(674.0,840.0){\rule[-0.200pt]{0.400pt}{4.818pt}}
\put(865.0,82.0){\rule[-0.200pt]{0.400pt}{4.818pt}}
\put(865,41){\makebox(0,0){40}}
\put(865.0,840.0){\rule[-0.200pt]{0.400pt}{4.818pt}}
\put(1056.0,82.0){\rule[-0.200pt]{0.400pt}{4.818pt}}
\put(1056,41){\makebox(0,0){50}}
\put(1056.0,840.0){\rule[-0.200pt]{0.400pt}{4.818pt}}
\put(1248.0,82.0){\rule[-0.200pt]{0.400pt}{4.818pt}}
\put(1248,41){\makebox(0,0){60}}
\put(1248.0,840.0){\rule[-0.200pt]{0.400pt}{4.818pt}}
\put(1439.0,82.0){\rule[-0.200pt]{0.400pt}{4.818pt}}
\put(1439,41){\makebox(0,0){70}}
\put(1439.0,840.0){\rule[-0.200pt]{0.400pt}{4.818pt}}
\put(100.0,82.0){\rule[-0.200pt]{322.565pt}{0.400pt}}
\put(1439.0,82.0){\rule[-0.200pt]{0.400pt}{187.420pt}}
\put(100.0,860.0){\rule[-0.200pt]{322.565pt}{0.400pt}}
\put(100.0,82.0){\rule[-0.200pt]{0.400pt}{187.420pt}}
\put(1279,820){\makebox(0,0)[r]{napier-0080aaa}}
\put(1299.0,820.0){\rule[-0.200pt]{24.090pt}{0.400pt}}
\put(119,113){\usebox{\plotpoint}}
\multiput(138.58,113.00)(0.498,0.784){75}{\rule{0.120pt}{0.726pt}}
\multiput(137.17,113.00)(39.000,59.494){2}{\rule{0.400pt}{0.363pt}}
\multiput(177.58,174.00)(0.499,0.784){149}{\rule{0.120pt}{0.726pt}}
\multiput(176.17,174.00)(76.000,117.492){2}{\rule{0.400pt}{0.363pt}}
\multiput(253.58,293.00)(0.499,0.754){151}{\rule{0.120pt}{0.703pt}}
\multiput(252.17,293.00)(77.000,114.542){2}{\rule{0.400pt}{0.351pt}}
\multiput(330.58,409.00)(0.499,0.632){149}{\rule{0.120pt}{0.605pt}}
\multiput(329.17,409.00)(76.000,94.744){2}{\rule{0.400pt}{0.303pt}}
\multiput(406.58,505.00)(0.499,0.650){303}{\rule{0.120pt}{0.620pt}}
\multiput(405.17,505.00)(153.000,197.713){2}{\rule{0.400pt}{0.310pt}}
\put(119.0,113.0){\rule[-0.200pt]{4.577pt}{0.400pt}}
\put(1279,779){\makebox(0,0)[r]{napier-0186aah}}
\multiput(1299,779)(20.756,0.000){5}{\usebox{\plotpoint}}
\put(1399,779){\usebox{\plotpoint}}
\put(119,113){\usebox{\plotpoint}}
\put(119.00,113.00){\usebox{\plotpoint}}
\multiput(138,113)(11.312,17.402){4}{\usebox{\plotpoint}}
\multiput(177,173)(10.846,17.696){7}{\usebox{\plotpoint}}
\multiput(253,297)(11.143,17.511){7}{\usebox{\plotpoint}}
\multiput(330,418)(11.040,17.576){7}{\usebox{\plotpoint}}
\multiput(406,539)(11.461,17.304){13}{\usebox{\plotpoint}}
\put(559,770){\usebox{\plotpoint}}
\sbox{\plotpoint}{\rule[-0.400pt]{0.800pt}{0.800pt}}%
\put(1279,738){\makebox(0,0)[r]{APAC NF-0080aaa}}
\put(1299.0,738.0){\rule[-0.400pt]{24.090pt}{0.800pt}}
\put(119,113){\usebox{\plotpoint}}
\multiput(139.41,113.00)(0.503,0.772){71}{\rule{0.121pt}{1.431pt}}
\multiput(136.34,113.00)(39.000,57.030){2}{\rule{0.800pt}{0.715pt}}
\multiput(178.41,173.00)(0.501,0.745){145}{\rule{0.121pt}{1.389pt}}
\multiput(175.34,173.00)(76.000,110.116){2}{\rule{0.800pt}{0.695pt}}
\multiput(254.41,286.00)(0.501,0.689){147}{\rule{0.121pt}{1.301pt}}
\multiput(251.34,286.00)(77.000,103.299){2}{\rule{0.800pt}{0.651pt}}
\multiput(330.00,393.41)(0.542,0.501){133}{\rule{1.069pt}{0.121pt}}
\multiput(330.00,390.34)(73.782,70.000){2}{\rule{0.534pt}{0.800pt}}
\multiput(407.41,462.00)(0.501,0.549){299}{\rule{0.121pt}{1.078pt}}
\multiput(404.34,462.00)(153.000,165.762){2}{\rule{0.800pt}{0.539pt}}
\multiput(559.00,631.41)(1.455,0.502){99}{\rule{2.509pt}{0.121pt}}
\multiput(559.00,628.34)(147.792,53.000){2}{\rule{1.255pt}{0.800pt}}
\multiput(712.00,684.41)(2.673,0.501){223}{\rule{4.457pt}{0.121pt}}
\multiput(712.00,681.34)(602.748,115.000){2}{\rule{2.229pt}{0.800pt}}
\put(119.0,113.0){\rule[-0.400pt]{4.577pt}{0.800pt}}
\sbox{\plotpoint}{\rule[-0.500pt]{1.000pt}{1.000pt}}%
\put(1279,697){\makebox(0,0)[r]{linear}}
\multiput(1299,697)(20.756,0.000){5}{\usebox{\plotpoint}}
\put(1399,697){\usebox{\plotpoint}}
\put(119,113){\usebox{\plotpoint}}
\multiput(119,113)(10.679,17.798){2}{\usebox{\plotpoint}}
\put(140.36,148.60){\usebox{\plotpoint}}
\put(151.51,166.10){\usebox{\plotpoint}}
\put(162.69,183.59){\usebox{\plotpoint}}
\put(173.56,201.26){\usebox{\plotpoint}}
\put(184.24,219.06){\usebox{\plotpoint}}
\put(194.92,236.86){\usebox{\plotpoint}}
\put(205.69,254.60){\usebox{\plotpoint}}
\multiput(217,272)(11.083,17.549){2}{\usebox{\plotpoint}}
\put(238.80,307.33){\usebox{\plotpoint}}
\put(249.47,325.12){\usebox{\plotpoint}}
\put(260.15,342.92){\usebox{\plotpoint}}
\put(270.83,360.72){\usebox{\plotpoint}}
\put(281.95,378.23){\usebox{\plotpoint}}
\put(293.34,395.57){\usebox{\plotpoint}}
\put(304.02,413.37){\usebox{\plotpoint}}
\multiput(314,430)(10.679,17.798){2}{\usebox{\plotpoint}}
\put(336.06,466.76){\usebox{\plotpoint}}
\put(347.07,484.36){\usebox{\plotpoint}}
\put(358.32,501.80){\usebox{\plotpoint}}
\put(369.26,519.43){\usebox{\plotpoint}}
\put(379.94,537.23){\usebox{\plotpoint}}
\put(390.62,555.03){\usebox{\plotpoint}}
\put(401.38,572.77){\usebox{\plotpoint}}
\put(412.41,590.35){\usebox{\plotpoint}}
\multiput(423,608)(11.312,17.402){2}{\usebox{\plotpoint}}
\put(445.18,643.29){\usebox{\plotpoint}}
\put(455.85,661.09){\usebox{\plotpoint}}
\put(466.78,678.74){\usebox{\plotpoint}}
\put(477.65,696.42){\usebox{\plotpoint}}
\put(488.33,714.21){\usebox{\plotpoint}}
\put(499.19,731.90){\usebox{\plotpoint}}
\put(510.41,749.36){\usebox{\plotpoint}}
\multiput(521,767)(11.083,17.549){2}{\usebox{\plotpoint}}
\put(542.89,802.48){\usebox{\plotpoint}}
\put(553.57,820.28){\usebox{\plotpoint}}
\put(564.67,837.81){\usebox{\plotpoint}}
\put(575.45,855.54){\usebox{\plotpoint}}
\put(578,860){\usebox{\plotpoint}}
\end{picture}
\caption{Speedup curves for Eco-Tierra single site entropy jobs. Two
organisms (0080aaa and 0186aah) were run on Napier, and one (0080aaa)
on the APAC National Facility.}
\label{eco-tierra-scaling}
\end{figure}

At the time of writing, Classdesc has only been deployed in one
production code (Eco-Tierra) \cite{Standish99a,Standish03a}. Figure
\ref{eco-tierra-scaling} shows the speedup curves on two machines:
{\em Napier}, a 28 processor SGI Power Challenge, which is now a
rather dated SMP architecture, and the {\em APAC National Facility}, a
Compaq SC cluster, consisting of 4 processor nodes coupled by a
Quadrics switch.

0186aah is a larger job than 0080aaa, so has more parallelism. It is
still scaling linearly at 24 processors on Napier. The constant offset
from linear scaling is due to the fact that the master thread does
very little, if any computation. For some reason, the 0186aah job is
not running correctly on the APAC NF, so no benchmark results are
available.

With 0080aaa, good scaling is obtained up to 24 processors, but on the
APAC NF performance falls off after that. This is most likely an
effect of the master process being unable to keep slaves
busy. However, the scaling discrepancy between Napier and the APAC NF
at lower process numbers indicates that network performance is
partially limiting. It should be borne in mind though that at 24
processes, the APAC NF is around 6 times faster than Napier!

\section{Discussion}

Classdesc was originally developed as part of the \EcoLab{} package to
allow it to handle arbitrary agent-based models\cite{Standish01b}. It
has been used in a project studying combinatorial
spacetimes\cite{Madina02}, and in several simulations using Graphcode, a 
C++ framework for agent-based simulation. Graphcode considers agents to 
be the vertices of arbitrary hypergraphs and is equipped with a graph 
partitioner to allow for the efficient execution of agent-based 
simulations on parallel computers. To achieve this, Graphcode uses 
Classdesc internally, so that objects may be serialized for transmission 
between cooperating processes. ClassdescMP is a natural outgrowth of 
Classdesc, and has been successfully employed in the Eco-Tierra 
project\cite{Standish99a,Standish03a}.

ClassdescMP is distributed as part of the Classdesc package, available
from http://parallel.hpc.unsw.edu.au/classdesc. It should compile and
be usable with any ANSI standard C++ compiler. It has been tested on a
variety of operating systems, including most Unices, Windows under
Cygwin and Mac OS X.

\section*{Acknowledgements}

The development of Classdesc and ClassdescMP was made possible with a
grant from the Australian Partnership of Advanced Computing. Benchmark
results were obtained on both APAC and ac3 facilities.


\end{document}